\documentclass[prb,showpacs,twocolumn,amsmath,amssymb]{revtex4}
\usepackage{graphicx}
\usepackage{dcolumn}
\usepackage{bm}
\begin{document}
\par

\title{Properties of mesoscopic superconducting thin-film
rings. London approach}
\author{V. G. Kogan, John R. Clem}
\affiliation{Ames Laboratory - DOE and Department of Physics and
Astronomy, Iowa State University, Ames Iowa 50011 }
\author{R. G. Mints}
\affiliation{School of Physics and Astronomy, Raymond and Beverly Sackler
Faculty of Exact Sciences, Tel Aviv University, Tel Aviv 69978, Israel}
\date{\today}
\begin{abstract}
Superconducting thin-film rings smaller than the film penetration depth
(the Pearl length) are considered. The current distribution, magnetic
moment, and  thermodynamic potential ${\cal F}(H,N,v)$ for a flat,
washer-shaped annular ring in a uniform applied field $H$ perpendicular
to the film are solved analytically within the London approach for a
state with winding number $N$ and a vortex at radius $v$ between the
inner and outer radii.
\end{abstract}
\maketitle
\section{Introduction}
Small flat rings made of thin superconducting films are of interest for
a variety of mesoscopic experiments.\cite{Davidovic 1,Davidovic 2}
SQUID-type devices is an example; another one is the study of fluxoid
dynamics in rings, as discussed by Kirtley {\it et al.}.\cite{John's
ring} The basic physics of such rings is governed by flux quantization,
as in the Little-Parks experiment; see, e.g., Ref.~\onlinecite{Tinkham}
or later publications \cite{Buzdin,Berger,Palacios,Sweigert} based on
the Ginzburg-Landau theory.
\par

There is a resurgence of experimental interest in this subject: The
interacting dipole moments in an array of superconducting rings provide
a model system for magnetism in Ising antiferromagnets. \cite{Davidovic
1,Davidovic 2} Moreover, there is  considerable interest in quantum
coherence effects in superconducting rings and their arrays for
potential applications in quantum computing.\cite{Ioffe,Mooij,Van der
Wal}
\par

Many quantitative details specific to the thin-film geometry can be
treated within the London approach, which is not bound by the strict
temperature restriction of Ginzburg-Landau models. In the thin-film
limit, for which the London penetration depth $\lambda$ obeys $\lambda
\gg d$, the film thickness, the fields and currents are governed by the
Pearl length $\Lambda=2\lambda^2/d$.~\cite{Pearl64} As we show below,
when the inner and outer radii $a$ and $b$ of the annular ring are
smaller than $\Lambda$, it is possible to obtain analytic solutions for
the energy of the ring in a uniform magnetic field $H$ applied
perpendicular to the ring plane with a vortex sitting in an arbitrary
position at the annular region between $a$ and $b$. The motion of
vortices between $a$ and $b$ provides the means for the ring to switch
between discrete states with different winding numbers $N$ defined
below. Study of these transitions is relevant for understanding the
telegraph noise observed in multiply connected mesoscopic
superconducting devices in general and in thin-film rings in
particular.\cite{John's ring}
\par

\section{Current distribution}
Let us consider a small thin-film ring of thickness $d\ll\lambda$
situated in the plane $z=0$ with inner and outer radii $a$ and $b$,
where $b$ is much smaller than the Pearl length $\Lambda=2\lambda^2/d$.
The ring is in a uniform applied field $H$ perpendicular to the film
plane; it may contain a vortex (or antivortex), the position of which
can be taken as $x=v,\,\, y=0$ (with the origin at the ring center and
$a<v<b$). The London equations for the local magnetic field ${\bm b}$
in the film interior read\cite{de Gennes}
\begin{equation}
{\bm b}+\frac{4\pi\lambda^2}{c} {\rm curl} {\bm j} =\pm \phi_0 {\hat
{\bm z}}\delta (x-v,y)\,.
\label{e1}
\end{equation}
Here, ${\bm j}$ is the current density$, {\hat{\bm z}}$ is the unit
vector perpendicular to the film plane, and $\phi_0=\pi\hbar c/|e|$ is
the value of the flux quantum. The upper sign holds for a vortex
whereas the lower one is for an antivortex, the convention retained
throughout this paper. Averaging this over the thickness $d$, one
obtains
\begin{equation}
b_z+\frac{2\pi\Lambda}{c} {\rm curl}_z {\bm g} =\pm \phi_0 \delta
({\bm r}-{\bm v}) ,
\label{e2}
\end{equation}
where ${\bm g}({\bm r})$ is the sheet current density, ${\bm r}=(x,y)$,
and ${\bm v}=(v,0)$. Equations (\ref{e1}) and (\ref{e2}) are valid
everywhere in the film except within a distance of the order of the
coherence length $\xi$ (vortex core) from the vortex or antivortex
axis, where the London equation (\ref{e1}) no longer holds.
\par

The distribution ${\bm g}({\bm r})$ can be found by solving Eq.\
(\ref{e2}), combined with the continuity equation and the Biot-Savart
integral that relates the field $b_z$ to the surface current:
\begin{equation}
{\rm div} {\bm g}=0,\quad
b_z({\bm r})=\int [{\bm g}({\bm r}')\times {\bm
R}/cR^3]_zd^2 {\bm r}' +H,
\label{e3}
\end{equation}
where ${\bm R}={\bm r}-{\bm r}'$.
\par

\subsection{The stream function}
In principle, Eqs.\ (\ref{e2}) and (\ref{e3}) determine the current
distribution. To solve these equations for the general case is a
difficult task, even for a disk.\cite{Fetter} However, for small
samples, as in our case for which $b\ll\Lambda $, the problem can be
solved.\cite{Kogan} To this end, let us introduce a scalar stream
function $G({\bm r})$, such that
\begin{equation}
{\bm g}={\rm curl}(G{\hat {\bm z}}).
\label{gg}
\end{equation}
The first of Eqs.\ (\ref{e3}) is then satisfied. It is easily seen that
the contours $G(x,y)=$ const coincide with the current streamlines.
Substituting Eq.\ (\ref{gg}) into (\ref{e2}), we obtain:
\begin{equation}
\frac{2\pi\Lambda}{c}\nabla^2G=\mp \phi_0\delta({\bm r}-{\bm v})+b_z.
\label{G''}
\end{equation}
The radial component of the current at the ring edges must be zero; this
means that the values of $G(x,y)$ at the edges are constants:
\begin{equation}
G(r=a,\varphi)=G_a,\qquad G(r=b,\varphi)=G_b,
\label{bc}
\end{equation}
where $(r,\varphi)$ are polar coordinates. Since $g_{\varphi}=-\partial
G/\partial r$, the total counterclockwise current (in the $\varphi$
direction) around the ring is
\begin{equation}
I=G_a-G_b.
\label{I}
\end{equation}
\par

The self-field of the ring currents {\it within the ring} can be
estimated using the Biot-Savart law: $\int d^2{\bm r}g/cR^2\sim g/c$.
Substitution of $b_z\sim H+g/c$ into Eq.\ (\ref{e2}) reveals that the 
self-field can be disregarded because $R\ll \Lambda$; i.e., we can set
$b_z=H$ in Eq.\ (\ref{G''}):
\begin{equation}
\frac{2\pi\Lambda}{c}\nabla^2G=\mp\phi_0\delta({\bm r}-{\bm v})+H .
\label{Poisson}
\end{equation}
\par

Since this equation is linear, we can look for a solution of the form
$G=G_v+G_H$, such that $G_v$ satisfies
\begin{equation}
\nabla^2G_v=\mp\frac{c\phi_0}{2\pi\Lambda}\delta({\bm r}-{\bm v}) ,
\label{Gv}
\end{equation}
and
\begin{equation}
\nabla^2G_H= \frac{c}{2\pi\Lambda} H.
\label{GH}
\end{equation}
One can say that $G_v$ describes the currents due to the vortex or
antivortex, whereas
$G_H$ is the response to the applied field.
\par

The boundary conditions (\ref{bc}) are imposed on the sum $G_v+G_H$. It
is convenient to require that
\begin{equation}
G_v(a)= G_v(b)=0\,,
\label{bc_Gv}
\end{equation}
and
\begin{equation}
G_H(a)=G_a\,,\qquad  G_H(b)=G_b\,.
\label{bc_GH}
\end{equation}
\par

In a uniform field $H$, $G_H$ can be taken as cylindrically symmetric.
Aside from an unimportant additive constant,
\begin{equation}
G_H(r)= \frac{cH}{8\pi\Lambda}\, r^2 + G_0\ln \frac{r}{a},
\label{GH(r)}
\end{equation}
where the constant $G_0$ is expressed in terms of the total current
(\ref{I}):
\begin{equation}
I =  -\frac{cH(b^2-a^2)}{8\pi\Lambda} - G_0\ln \frac{b}{a}\,.
\label{IGH}
\end{equation}
To evaluate $I$ we use the London equation in the form
\begin{equation}
{\bm g}=-\frac{c\phi_0}{4\pi^2\Lambda}\Big(\nabla\theta
+\frac{2\pi}{\phi_0}{\bm A}\Big)\,.
\label{London}
\end{equation}
Here $\theta$ is the order parameter phase, the topology of which plays
a major role in our problem.
\par

When the ring is traversed around a circle of radius $r$ in the
positive direction of the azimuth $\varphi$, the phase changes by
$-2\pi N$, where $N$ is an integer which is commonly called the winding
number or the vorticity.  If there are no vortices in the annulus, the
integer $N$ is the same for any contour within the annulus and  we
consider the state as having winding number $N$. In other words, the
state of the system is characterized by the integer $N$ and the
continuous variable $H$. In zero field, $N>0$ corresponds to  positive
currents $g_{\varphi}$ and positive magnetic moments $\mu_z$.
\par

In the presence of a vortex, however, the situation is different. For
contours encircling the ring's hole, the winding number is $N$ at
contours that do not include the vortex position; the number is $N+1$
for those that do. The state is now characterized by the vortex
position ${\bm v}$ in addition to $N$ and $H$. In this paper, when  the
state of the ring with a vortex is characterized by the variables $N,
H,$ and $v$, it is implied that the integer $N$ describes the phase
topology on contours that do not include the vortex. The generalization
to antivortices is obvious.
\par

Coming back to evaluation of the total current, we integrate
$I=\int_a^bg_\varphi (r,\varphi)dr$ over $\varphi$ to get:
\begin{equation}
I = \int_a^b \bar {g}_\varphi (r) dr\,,
\end{equation}
where the azimuthal average of
$g_{\varphi}(r,\varphi)$ is
\begin{equation}
\bar{g}_\varphi(r)=\int_0^{2\pi}g_{\varphi}\frac{d\varphi}{2 \pi}=
\frac{c\phi_0}{4\pi^2 \Lambda}\left(\frac{N}{r}
-\frac{\pi H}{\phi_0}r\right)\,,
\label{gbar}
\end{equation}
for $r<v$, and
\begin{equation}
\bar{g}_\varphi(r>v)=
\frac{c\phi_0}{4\pi^2 \Lambda}\left(\frac{N\pm 1}{r}
-\frac{\pi H}{\phi_0}r\right) . \label{gbar1}
\end{equation}
We now readily evaluate the total current and the constant $G_0$:
\begin{equation}
G_0=-\frac{c\phi_0}{4\pi^2 \Lambda}\Big[N
\pm \frac{\ln(b/v)}{\ln(b/a)}\Big] .
\label{G_0general}
\end{equation}
\par

Note that $G_0$, which determines the part $G_H$ of the stream
function, depends on whether or not a vortex is present at the ring and
on its position. In other words, the current distribution is not a
simple superposition of the currents generated by a vortex in zero
field and of those existing in the state $(N,H)$ in the vortex absence.
\par

\subsection{Electrostatic analogy and exact solution}
To find the solution of Eq. (\ref{Gv}) for the vortex-generated stream
function $G_v$ subject to the boundary conditions (\ref{bc_Gv}), we
observe that the problem  is  equivalent to the two-dimensional one for
the electrostatic potential generated by a line charge $\pm
c\phi_0/8\pi^2\Lambda$ at the point ${\bm v}$ situated between two
coaxial grounded metallic cylinders with radii $a$ and $b$. The
necessary conformal mapping procedure is given   in
Ref.~\onlinecite{Morse}:
\begin{eqnarray}
G_v(w,v)=\pm\frac{c\phi_0}{4\pi^2\Lambda}{\rm
Re}\left[\ln\frac{A(w,v)}
{B(v)}\right],\quad w=re^{i\varphi},\nonumber\\
A(w,v)=\frac{{\rm cn}[2\gamma\ln(v/a),m]} {{\rm
sn}[2\gamma\ln(v/a),m]}- \frac{{\rm cn}[\gamma\ln(v/w),m]}{{\rm
sn}[\gamma\ln(v/w),m]},
\label{morse}\\
B(v)=\frac{{\rm cn}[2\gamma\ln(v/a),m]}{{\rm sn}[2\gamma\ln(v/a),m]}-
\frac{{\rm cn}[\gamma\ln(v/a),m]}{{\rm
sn}[\gamma\ln(v/a),m]},\nonumber
\end{eqnarray}
where ${\rm cn}(x),{\rm sn}(x)$ are the Jacobi elliptic functions,
$\gamma={\rm K}(m)/\ln(b/a)$ with ${\rm K}(m)$ being the complete
elliptic integral (in the notation of Ref.~\onlinecite{Abr}), and the
parameter $m$ is chosen to satisfy
\begin{equation}
{\rm K}(1-m)\ln(b/a)=\pi{\rm K}(m).
\label{eq15}
\end{equation}
As an example, we find that for $b/a=2$ (such rings were studied in
Ref.~\onlinecite{John's ring}), $m= 1.048\times 10^{-5}$, and
$\gamma=2.266$.
\par

For rings with $1<b/a<2$, we have $0<\ln(b/a)<1$. Solving Eq.\
(\ref{eq15}) numerically, one can see that $m\ll 1$. Expanding
functions ${\rm K}(m)$ and ${\rm K}(1-m)$ for small $m$, one obtains
with a high accuracy:
\begin{equation}
m=16\exp\left[-\frac{\pi^2}{\ln(b/a)} \right],\qquad
\gamma=\frac{\pi}{2\ln(b/a)} .
\label{eq16}
\end{equation}
One can now set $m=0$ in Eq.\ (\ref{morse}) to obtain
\begin{equation}
G_v \approx\pm\frac{c\phi_0}{4\pi^2\Lambda} {\rm Re}\left\{
\ln\frac{\sin[\pi\ln (vw/a^2)/2\ln(b/a)]}{\sin[\pi\ln
(v/w)/\ln(b/a)]}\right\}.
\label{approxG}
\end{equation}
\par

At the vortex  position $r=v,\quad \varphi =0$, this function diverges
logarithmically. One can find $ G_v({\bm v})$ [and the self-energy
$\epsilon_v= \phi_0 |G_v(v)|/2c$] by introducing the standard cutoff at
a distance $\xi$ from the vortex   axis at ${\bm v}$:
\begin{equation}
\epsilon_v \approx\frac{\phi_0^2}{8\pi^2\Lambda}\ln
\left[\frac{2v\ln(b/a)}{\pi\xi}
\sin\frac{\pi\ln(v/a)}{\ln(b/a)}\right].
\label{e_v}
\end{equation}
The relative difference between this expression and the exact energy is
less than $2\times 10^{-12}$ for $b/a\le 2$.\cite{rem1} Formally, the
logarithmic factor in Eq.\ (\ref{e_v}) goes to $-\infty$ if $v\to a$ or
$v\to b$. However, this expression fails when the vortex   is within
roughly $\xi$ of the inner or outer radius. In fact, $\epsilon_v$ of
Eq.\ (\ref{e_v}) becomes equal to zero at $v_a=a(1+\xi/2a)$ and
$v_b=b(1-\xi/2a)$.
\par

\begin{figure}
\hbox{
\includegraphics[width=0.475\hsize ]{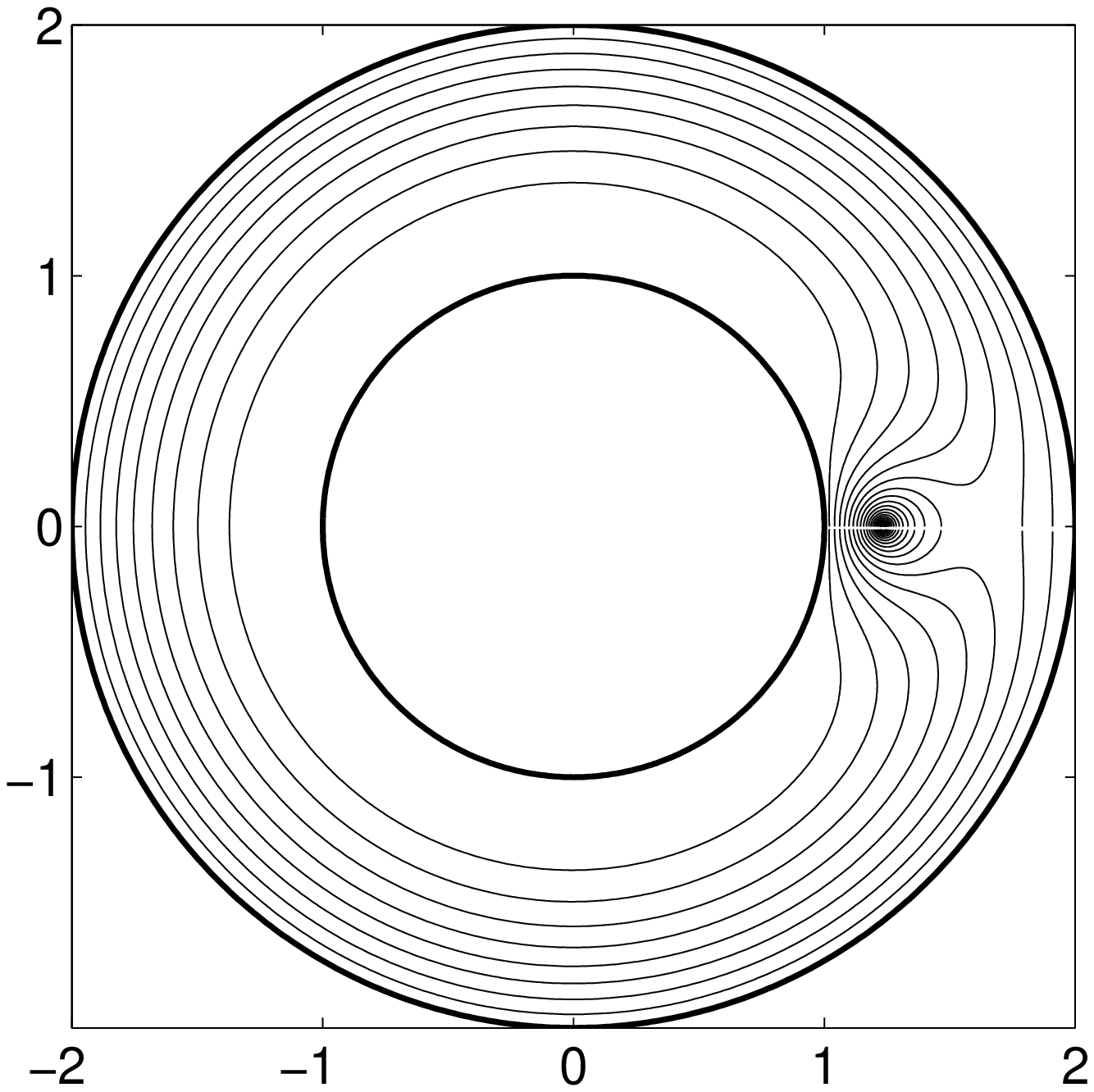}
\includegraphics[width=0.475\hsize]{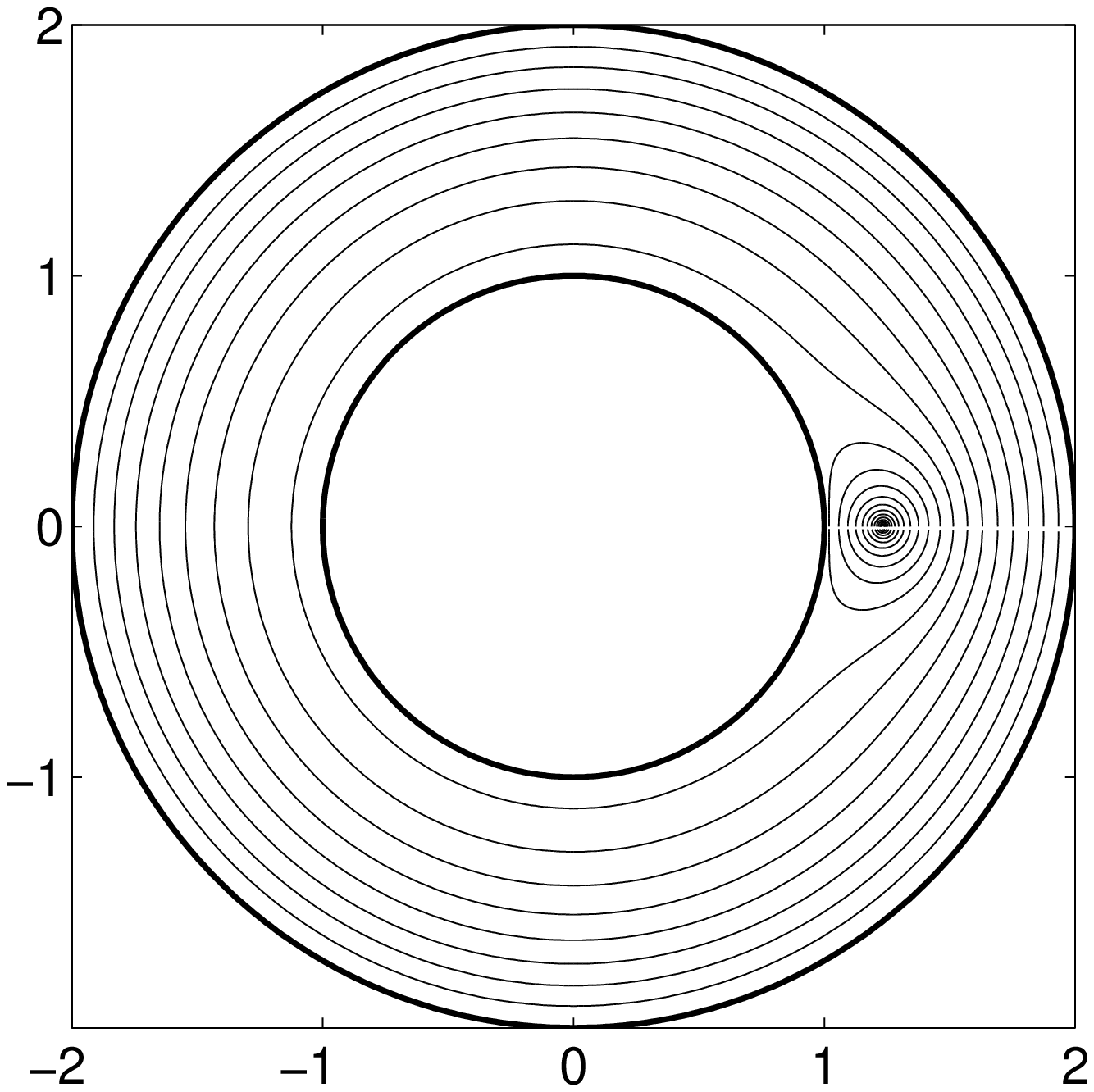}}
\caption{Current streamlines in a ring with $b/a=2$, $N=2$, and vortex
or antivortex position $v/a=1.2$ in the normalized field $h=H/H_0=2.3$
[see Eq.\ (\ref{H0})]. The streamlines are plotted as contours of
$G=G_v+G_H={\rm const}$. The left panel is for a vortex; the
streamlines shown that encircle the hole have clockwise flow, whereas
the flow around the vortex is counterclockwise. The right panel shows
an antivortex; all streamlines shown have clockwise flow.}
\label{f1}
\end{figure}

Thus, the problem of the current distribution in the ring is solved:
the vortex-generated part $G_v$ of the stream function is given in Eqs.\
(\ref{morse}), while  $G_H$ is determined by Eqs.\ (\ref{GH(r)}) and
(\ref{G_0general}). The current streamlines are given by the contours
$G(r,\varphi)= G_v(r,\varphi) + G_H(r) =$ const; Fig.
\ref{f1} shows examples of current streamlines for a vortex and an
antivortex.
\par

It is worth mentioning that the same method based on application of
conformal mapping to problems of the two-dimensional electrostatics can
be utilized to obtain current distributions for $n$ vortices equally
spaced along a circle of radius $a<v<b$ of the ring.\cite{Morse}
\par

\par

\section{Energy}
\subsection{Zero applied field}

Let us first consider the energy $E(N,v)$ of the ring with a vortex at
$v$ in the state $N$ in zero applied field. This energy is a sum of the
magnetic and kinetic contributions. The magnetic part is\cite{Landau}
\begin{eqnarray}
E_m &=&\frac{1}{2c}\int {\bm A}\cdot{\bm
g}\, d^2{\bm r}\nonumber\\&=&-\frac{1}{2c}\int \left(\frac
{2\pi\Lambda}{c}{\bm g}+ \frac {\phi_0}{2\pi}
\nabla\theta\right)\cdot{\bm g}\,d^2{\bm r},\label{Em}
\end{eqnarray}
where we use Eq.\ (\ref{London}) to express the vector potential ${\bm
A}$ at the ring in terms of ${\bm g}$ and the phase $\theta$. The
supercurrent kinetic energy $E_k$ is the integral over the film volume
of the quantity $2\pi\lambda_L^2j^2/c^2=\pi\Lambda g^2/c^2d$.\cite{de
Gennes}  We find readily that this energy is equal in value and
opposite in sign to the term containing $g^2$ in Eq.\ (\ref{Em}), so
that
\begin{equation}
E(N,v) \equiv E_{m} + E_k= -\frac{\phi_0}{4 \pi c} \int  (\nabla \theta
\cdot {\bm g})d^2
{\bm r}.
\label{E_tot}
\end{equation}
\par
The integrand here can be further transformed in terms of the stream
function $G$:
\begin{equation}
\nabla\theta \cdot {\rm curl}(G{\bm z})={\rm div} (G{\bm z} \times
\nabla\theta )+G{\bm z}\cdot{\rm curl}\nabla\theta.
\label{transform}
\end{equation}
Substituting this into Eq.\ (\ref{E_tot}), we use Gauss's theorem to
evaluate the contribution of the first term:
\begin{eqnarray}
& &\int {\rm div} (G{\bm z} \times
\nabla\theta ) d^2{\bm r}
\nonumber\\
&=&\int_0^{2\pi}\left[-G(b,\varphi)\frac{\partial\theta}
{\partial\varphi} \Big|_b
+G(a,\varphi)\frac{\partial\theta}{\partial\varphi}
\Big|_a\right]d\varphi
\label{gauss} \\
&=&2\pi [(N\pm 1)G_b-NG_a] .\nonumber
\end{eqnarray}
Here we have used the boundary conditions (\ref{bc}) and the   phase
change is $-2\pi N$ upon circling the inner radius and $-2\pi (N\pm 1)$
the outer radius, provided a single vortex (antivortex) is present in the
annulus.
\par

Integrating the second term of Eq.\ (\ref{transform}), we use the basic
quantization property of the phase [curl$_z\nabla\theta =\mp 2\pi\delta
({\bm r}-{\bm v})$], take the curl of Eq.(\ref{London}), and compare
the result with Eq.\ (\ref{e2}). Then, we obtain the magnetic and
kinetic energy of the persistent currents in the ring:
\begin{equation}
E(N,v)= \frac{\phi_0}{2c}[G(v)-(N\pm 1)G_b+NG_a].
\label{E0(N,v)}
\end{equation}
Note that $G(v)=G_v(v)+G_H(v)$, and the vortex self-energy is
$\epsilon_v= \phi_0 |G_v(v)|/2\pi$  (see Ref.~\onlinecite{Kogan}).
Utilizing Eqs.\ (\ref{GH(r)}) and (\ref{G_0general}), we obtain:
\begin{equation}
E(N,v)=\epsilon_v(v) + \epsilon_0 \Big[N\pm \frac{\ln(b/v)}{\ln(b/a)
}\Big]^2 ,
\label{E(N,v)}
\end{equation}
where we introduce the energy scale
\begin{equation}
\epsilon_0=\frac{\phi_0^2\ln(b/a)}{8\pi^2\Lambda} .\label{eps_0}
\end{equation}
\par

If $v=b$, i.e., if there is no vortex in the annulus,  we have
\begin{equation}
E_0  = \frac{\phi_0}{2c}(G_a -G_b)N =
\frac{\phi_0}{2c}NI.
\label{E_1}
\end{equation}
\par

\subsection{Magnetic moment}
By definition, the $z$ directed magnetic moment reads:\cite{Landau}
\begin{equation}
\mu =\frac{1}{2c}\int_a^bdr\,r^2 \int_0^{2\pi}d\varphi
g_\varphi(r,\varphi )\,=\frac{\pi}{c}\int_a^br^2 \bar{g}_\varphi(r )dr.
\label{mu1}
\end{equation}
This is easily evaluated using Eqs.\ (\ref{gbar}), (\ref{gbar1}):
\begin{equation}
\mu =\frac{\phi_0(b^2-a^2)}{8\pi\Lambda}\Big(N\pm\frac{b^2-v^2}
{b^2-a^2}\Big) -\frac{ H}{ 16\Lambda}(b^4-a^4).
\label{mu+v}
\end{equation}
It is worth noting that the magnetic moment is evaluated here without
need to employ the explicit current distribution which could have
been obtained from the known stream function $G_v+G_H$.
\par                     

Setting $v=b$, we obtain the moment $\mu_0$ in the ``pure" state $N$
with no vortex at the ring:
\begin{equation}
\mu_0(N,H)  =\frac{\phi_0(b^2-a^2)}{8\pi\Lambda}N -\frac{ H}{ 16\Lambda}
(b^4-a^4).
\label{mu-v}
\end{equation}
The result can be verified by direct calculation of $\mu$ using the
current distribution in the absence of  vortices. If we set $v=a$ in
(\ref{mu+v}), we obtain
$\mu_0 (N\pm 1,H)$.
\par

It is instructive to observe that the exact moment (\ref{mu+v}) in the
presence of a vortex at $v$ can be considered as a sum of the magnetic
moment of a vortex-free ring  with inner and outer radii $a$ and $v$
having winding number $N$ and another ring  with inner and outer radii
$v$ and $b$ with vorticity $N+1$.
\par

\subsection{Free energy in field}
The thermodynamic potential ${\cal F}$ which is minimum in equilibrium
at a given applied field is defined by a differential relation
\begin{equation}
\delta {\cal F} = -{\bm \mu}\cdot \delta{\bm H}=-\mu\, \delta H\,,
\label{dF}
\end{equation}
where ${\bm \mu}(N,v,H)$ is the system magnetic moment.\cite{Landau}
Using Eq.\ (\ref{mu+v}), we readily obtain ${\cal F}$ by integrating
Eq.\ (\ref{dF}) over $H$ from $0$ to $H$:
\begin{eqnarray}
{\cal F}_{\pm} = E(N,v) &-& \frac{\phi_0H(b^2-a^2)}{8\pi\Lambda}
\left(N\pm\frac{b^2-v^2} {b^2-a^2}\right)\nonumber\\ &+&\frac{H^2}{
32\Lambda}(b^4-a^4),
\label{F1}
\end{eqnarray}
where the zero-field energy $E(N,v)$ is given in Eq.\ (\ref{E(N,v)}).
After simple algebra we arrive at our main result,
\begin{eqnarray}
{\cal F}_\pm(N,v,H)&=&\epsilon_v(v)  + \epsilon_0\Bigg[
\left(N\pm\frac{\ln(b/v)}{\ln(b/a)}\right)^2
\nonumber\\
&-&2h\left(N\pm\frac{b^2-v^2}{b^2-a^2}\right) + h^2\chi\Bigg],
\label{Gibbs1}
\end{eqnarray}
where we have introduced a dimensionless field $h=H/H_0$, where
\begin{equation}
H_0=\frac{2\phi_0\ln(b/a)}{\pi(b^2-a^2)}  ,
\label{H0}
\end{equation}
and a geometric factor $\chi$ given by
\begin{equation}
\chi =\frac{b^2/a^2+1}{b^2/a^2-1}\ln\frac{b}{a}.
\end{equation}
For a narrow ring $b/a-1=\eta\ll 1$, $\chi =1+{\cal O}(\eta^2)$. With
increasing $b/a$, $\chi$ grows slowly: for $b/a=2$, $\chi\approx
1.155$, and  $\chi\approx \ln(b/a)$ for large values of $b/a$.
\par

\subsection{Vortex-free state}
For the vortex-free state, Eq.\ (\ref{Gibbs1}) yields:
\begin{equation}
{\cal F}_0(N,h)  = \epsilon_0 \Big(N^2-2N h  + h^2 \chi \Big).
\label{pure-state}
\end{equation}
It is seen that  ${\cal F}_0(N_1,h)={\cal F}_0 (N_2,h)$ at the field
$h=(N_1+N_2)/2$ for any $N_1$ and $N_2$. In particular, for $N_1=N$ and
$N_2=N+1$, the energies $ {\cal F}_0(N,h)={\cal F}_0 (N+1,h)$ at $h=
N+1/2$. In other words, at this field the system  might be in either of
the states $N$ or $N+1$ having the same energy.
\par

One can readily check that the thermodynamic potential ${\cal
F}_0(N,h)$ of the vortex-free state coincides with the {\it kinetic}
energy of the supercurrents in the ring.
\par

A transition from a  state with vorticity $N$ to one with  vorticity
$N+1$ can  happen when a vortex, carrying  unit vorticity, enters at
the outer radius, crosses the ring, and annihilates at the inner
radius. Alternatively, such a transition can be accomplished, starting
with initial vorticity $N$, when an antivortex, carrying vorticity --1,
enters the annulus at the inner radius $a$, crosses the ring, and
annihilates at the outer radius $b$, leaving behind vorticity $N+1$
trapped in the ring.
\par

\begin{figure}
\includegraphics[width=0.95\hsize]{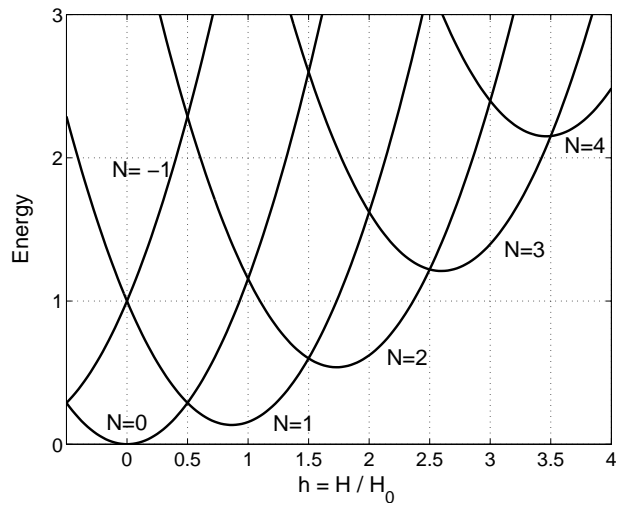}
\caption{The energy ${\cal F}_0$ of vortex-free states of a ring with
$b/a=2$ for vorticities $N$ from 0 to 4 versus applied field. The
energy and field units are defined in Eqs. (\ref{eps_0}) and
(\ref{H0}). Note that e.g. for $0.5<h<1.5$, $N=1$ corresponds to the
ground state. The nearest ``excited" state is $N=0$ for $0.5<h<1$,
whereas for $1<h<1.5$ it is $N=2$.}
\label{f2}
\end{figure}

The field dependence of ${\cal F}_0$ for $0\le N\le 4$ is shown in Fig.
\ref{f2}, which illustrates the above features. It shows that in fields
$N-1/2<h<N+1/2$, the minimum energy belongs to the state $N$. However,
the first excited state is $N-1$ for $N-1/2<h<N$, whereas it is $N+1$
for $N <h<N+1/2$. It should be also noted that in a given field $h$,
the {\it ground state} winding number is the integer nearest to $h$.
\par

Although the ground state vorticity $N$   changes with field,  the
energy {\it difference} between the lowest and the ``first excited
state" is, in fact, periodic in $h$ with the period $\Delta h =1$. It
is easy to check that   the difference ${\cal F}_0(N,h)-{\cal
F}_0(N-1,h )= {\cal F}_0(N+1,h+1)-{\cal F}_0(N,h+1)$. This fact has
implications for the transition probabilities from the ground state to
the nearest excited state.
\par

\par

It is worth observing that the energy (\ref{Gibbs1}) differs from that
of the vortex-free state, ${\cal F}_0$ of Eq.\ (\ref{pure-state}), by
the vortex self-energy $\epsilon_v(v) $ and by the addition to $N$ of the
$v$-dependent terms that vary from unity to zero as the
vortex moves from $a$ to $b$.
\par
\subsection{Potential barriers for vortices crossing the ring}
We are now in a position to evaluate the energy barrier that must be
overcome for a vortex or an antivortex to move between the inner and
outer radii of the ring. Since $G_v=0$ at both $a$ and $b$, we have at
$v=b$:
\begin{equation}
{\cal F}_+(N,b)={\cal F}_-(N,b)={\cal F}_0(N),
\label{FNb}
\end{equation}
where the energy ${\cal F}_0(N)$ of the vortex-free states is given by
Eq.\ (\ref{pure-state}); the subscript ``+" is for a vortex, whereas
``--" stands for an antivortex (the field argument $H$ is suppressed
for brevity). In other words, the addition of either a vortex or an
antivortex at the outer radius does not change the vorticity of the
system. On the other hand, at $v=a$ we have:
\begin{equation}
{\cal F}_+(N,a)={\cal F}_0(N+1),\quad {\cal F}_-(N,a)={\cal
F}_0(N-1);
\label{FNa}
\end{equation}
i.e., moving a vortex (antivortex) from $b$ to $a$ changes the
vorticity by $+1$ ($-1$), an obvious consequence of the system
topology.
\par

Starting from a vortex-free state $N$, the motion of a vortex between
$a$ and $b$ can change the system energy in two ways. If a vortex moves
{\it in}, starting from the outer radius $b$, the spatial dependence of
the potential energy barrier against vortex entry is given by the
difference
\begin{equation}
{\cal F}_+(N,v)-{\cal F}_0(N) =V_{in}^+(N,v) ,
\label{Vin+}
\end{equation}
so that $V_{in}^+(N,b)=0$, while $V_{in}^+(N,a)=\epsilon_0(2N+1-2h)$,
which corresponds to the energy of transition between ``pure" states
$N$ and $N+1$. On the other hand, if a vortex moves {\it out}, starting
from the inner radius $a$, the spatial dependence of the potential
energy barrier against vortex exit is given by the function
\begin{equation}
V_{out}^+(N,v)={\cal F}_+(N-1,v)-{\cal F}_0(N ) ,
\label{Vout+}
\end{equation}
such that $V_{out}^+(N,a)=0$, while
$V_{out}^+(N,b)=\epsilon_0(2h-2N+1)$, which  corresponds to the energy
of transition from the state $N$ to the state $N-1$.
\par

When $h=N+1/2$, we obtain $V_{in}^+(N,a)=V_{in}^+(N,b)=0$,
$V_{out}^+(N+1,b)=V_{out}^+(N+1,a)=0$ for any $N$. Moreover, the
potential barriers  $V_{in}^+(N,v)$ and $V_{out}^+(N+1,v)$ are
identical.
\par

Similarly, the motion of an antivortex between $a$ and $b$ can change
the system energy in two ways.  If an antivortex   moves {\it in},
starting from the outer radius $b$, the spatial dependence of the
potential energy barrier against antivortex entry is given by the
function
\begin{equation}
V_{in}^-(N,v)={\cal F}_-(N,v)-{\cal F}_0(N) ,
\label{Vin-}
\end{equation}
so that $V_{in}^-(N,b)=0$, while $V_{in}^-(N,a)=\epsilon_0(2h-2N+1)$,
which corresponds to the energy of transition from the state $N$ to the
state $N-1$. On the other hand, if an antivortex  moves {\it out},
starting from the inner radius $a$, the spatial dependence of the
potential energy barrier against vortex exit is given by the function
\begin{equation}
V_{out}^-(N,v)={\cal F}_-(N+1,v)-{\cal F}_0(N ) ,
\label{Vout-}
\end{equation}
such that $V_{out}^-(N,a)=0$, while $V_{out}^-(N,b)=\epsilon_0(2N+1-2h)$,
which corresponds to the energy of transition from the state $N$ to the
state $N+1$.
\par

Note that $V_{out}^-(N,v)$ differs from $V_{in}^-(N+1,v)$ only by a 
 constant: $V_{out}^-(N,v)-V_{in}^-(N+1,v) = {\cal F}_0(N
+1)-{\cal F}_0(N )=\epsilon_0(2N+1-2h)$. When  this constant is zero,
i.e. when $h=N+1/2$, the barriers $V_{out}^-(N-1,v)$ and
$V_{in}^-(N,v)$ become identical.
\par

We conclude the discussion of barriers by pointing out that  although
the total energies ${\cal F}_\pm(N,H,v)$ are quadratic in $N$ and $H$,
the barrier functions are linear in these variables. For example, we
have for a vortex entry at $b$:
\begin{eqnarray}
&&V_{in}^+(N,H,v)=\epsilon_v(v) \label{barrier+in}\\
&&+ \epsilon_0\left\{
\left[2N+\frac{\ln(b/v)}{\ln(b/a)}\right]\frac{\ln(b/v)}{\ln(b/a)}
-2h \frac{b^2-v^2}{b^2-a^2}  \right\}.\nonumber
\end{eqnarray}

\par

\section{Narrow rings}
The energy (\ref{Gibbs1}) simplifies for  narrow rings of width
$W=b-a\ll a$, i.e., for
\begin{equation}
\eta=W/a  << 1.
\label{eta}
\end{equation}
We obtain in the linear approximation in $\eta$:
\begin{eqnarray}
{\cal F}_{\pm}&=&\epsilon_v(z)+ \epsilon_0\,(N\pm z-h)^2 \,,\label{b=a}\\
\epsilon_v&=&{\frac{\phi_0^2}{8\pi^2\Lambda}}
\ln\Big(\frac{2a\eta}{\pi\xi}\sin\pi z\Big)\,,\quad
z=\frac{b-v}{b-a}\,.
\label{ev_narrow}
\end{eqnarray}
Here,   $0<z<1$; the energy and field scales are
\begin{equation}
\epsilon_0 \approx \frac{\phi_0^2}{8\pi^2\Lambda}\,\eta\,,\quad\
H_0 \approx \frac{\phi_0}{\pi a^2}\,.
\label{e0_narrow}
\end{equation}
In particular, we have for the vortex-free state ($z=0$):
\begin{equation}
{\cal F}_0\approx\frac{\phi_0^2\eta}{8\pi^2\Lambda}(N-h)^2=
\frac{\phi_0^2\eta}{8\pi^2\Lambda}
\left(N-\frac{\pi a^2H}{\phi_0}\right)^2,
\label{Fo}
\end{equation}
an expression similar to that for thin cylinders.\cite{Tinkham,Barone}
\par

The potential barriers  defined in Eqs. (\ref{Vin+})-(\ref{Vout-}) are
now easily evaluated. We focus here on $V_{in}^+$ for a vortex, which
reads in linear approximation in $\eta$:
\begin{equation}
V_{in}^+  \approx \epsilon_v+ \epsilon_0z\, (2N-2h+ z )\,.\label{Vin+z}
\end{equation}
One readily verifies that this potential reaches maximum at
$z_m=1/2+{\cal O}(\eta)$, i.e., at $ v_m =(b+a)/2$.
\par

\begin{figure}
\includegraphics[width=0.95\hsize]{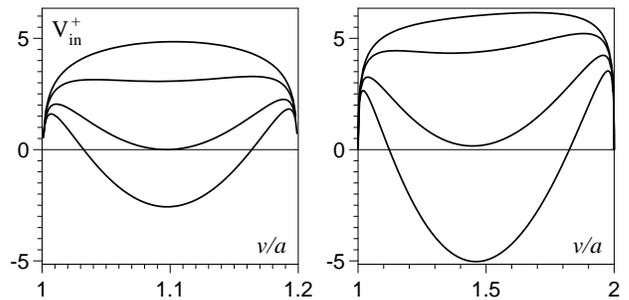}
\caption{The energy barriers $V_{in}^+(N,H,v)$ for the vortex entry at
the outer radius $b$  calculated numerically according to Eq. (\ref{Vin+}).
The energy
$V_{in}^+ $ is given in units of
$\phi^2/8\pi^2\Lambda$. The vortex position $v$ is normalized on the
inner radius $a$. The left panel: $b/a=1.2$; the parameters from top to
bottom are $h=N=3,110,295$, and $ 450$; $h$ is
normalized to $H_0$ of Eq. (\ref{H0}). The right panel:
$b/a=2$; the parameters from top to bottom are $h= N=3,10,28$, and
$50$. Note: the normalization field $H_0$ for this set differs from the
first.}
\label{fig3}
\end{figure}

To study the behavior of the potential $V_{in}^+(N,H,v)$ near its
maximum in increasing fields, one should go to higher order terms in
the small $\eta$. To avoid cumbersome algebra, we use the fact that in
the ground state $|h-N| \le 1/2$ so that in large fields we can set
$N\approx h$ and study the behavior of the function $V_{in}^+(h,h,v)$.
Of a particular interest is the curvature of the potential barrier at
$v=v_m$. Numerical experimentation shows that in large fields, the
maximum of $V_{in}^+$ is situated close to  $ v_m =(b+a)/2$.
Differentiating twice the function $V_{in}^+(h,h,v)$ with respect to
$v$ (this is easily done using Mathematica) and setting $v= v_m =
a+W/2$, we find that the curvature turns zero at
\begin{equation}
H\approx H_1 = \frac{\pi \phi_0}{4 W^2} \,.
\label{Hx}
\end{equation}
For $H>H_1$, the potential barrier $V_{in}^+(h,h,v_m)$ acquires a local
minimum. Examples are shown in Fig. \ref{fig3}. A vortex at this minimum
is in a {\it metastable} state provided $V_{in}^+(h,h,v_m)>0$.
\par

With increasing field, the depth of the minimum increases and at a
field  that may be called the low critical field, $H_{c1}$,
$V_{in}^+(h,h,v_m)$ becomes equal to zero. This is the minimum field at
which vortices can nucleate at the ring and stay there in  {\it stable}
equilibrium. We can estimate this field by setting
$V_{in}^+(h,h,v_m)=0$ in  Eq. (\ref{barrier+in}). For narrow rings we
obtain:
\begin{equation}
H_{c1} \approx \frac{2 \phi_0}{\pi W^2}\,\ln\frac{2W}{\pi\xi}\,.
\label{Hc1}
\end{equation}
\par

Clearly, with the further field increase, the  barriers on both sides
of the minimum are suppressed, while the points $v_0$ where
$V_{in}^+=0$ are pushed towards the annulus edges at $a$ and $b$. The
critical field $H^*$ at which the ``edge barrier" near $b$ disappears
can be estimated by setting $V_{in}^+(h,h,v_0)=0$ in Eq. (\ref{barrier+in})
and considering the limit $v_0\approx b-\xi$. After straightforward
algebra we obtain:
\begin{equation}
H^* \approx \frac{  \phi_0\ln 2}{\pi\,\xi\, b}\,\,
\frac{\ln(b/a)}{2\ln(b/a)-1+a^2/b^2}\,.
\label{Hc}
\end{equation}
For narrow rings, this reduces to
\begin{equation}
H^* \approx \frac{  \phi_0\ln 2}{2\pi\,\xi\, W}\,,
\label{Hc_narrow}
\end{equation}
whereas for rings with a small hole, $a\ll b$,
\begin{equation}
H^* \approx \frac{  \phi_0\ln 2}{2\pi\,\xi\,
b}\,\left[1+\frac{1}{2\ln(b/a)}\right]\,.
\label{Hc_pinhole}
\end{equation}
\par

It is of interest to note that the field $H_1$ of Eq. (\ref{Hx}) is
temperature independent, $H_{c1}$ of Eq. (\ref{Hc1}) is only weakly
depends on $T$, whereas $H^*$ vanishes as $T\to T_c$ as
$\sqrt{1-T/T_c}$. The last two are to be compared with the bulk
$H_{c1}$ and $H^*$, the bulk characteristics of the Bean-Livingston
barrier, both of which are linear in $1-T/T_c$ near $T_c$.\cite{de
Gennes}
\par

\section{Discussion}

The behavior of narrow rings is closely related to that of long strips.
The field given in Eq.\ (\ref{Hx}) was derived previously in a study of
long strips of width $W$,  where $H_1$ was called the {\it vortex
exclusion field}.~\cite{Clemunpub}  According to this theory, when a
superconducting strip is cooled through $T_c$ in an ambient magnetic
field $H_a$, vortices should be  excluded from the strip when $H_a$ is
less than $H_1$, because the free-energy  then has a global maximum at
the strip center (at $v_m$ in our notation). On the other hand, when
$H_a$ is greater than $H_1$, there is a local free-energy minimum at
$v_m$, where the probability   of finding a vortex is proportional to
$\exp(-V_{min}^+/k_BT)$. Near $T_c$, this probability is close to
unity. However, as $T$ decreases, the characteristic energy scale,
$\phi_0^2/8\pi^2\Lambda(T)$, increases rapidly.  The vortices are then
``frozen in"  a potential well similar to those shown in Fig.
\ref{fig3} at a characteristic freeze-in temperature $T_f$,  which for
most superconductors is estimated as being very close to $T_c$. The
value of $\Lambda(T_f)$ turns out larger than strip widths of the order
of a few $\mu$m, which justifies calculations done in the thin-film
limit. The above arguments suggest that SQUIDs to be cooled and
operated in the earth's magnetic field should be made entirely of
narrow lines in order to avoid flux noise due to thermal agitation of
vortices trapped in the lines.  Experiments by Dantsker {\it et
al.}~\cite{Dantsker97} are consistent with this conclusion.
\par

The solutions presented here for  currents and energies in thin
mesoscopic superconducting rings are of importance for the physics of
isolated rings (such as in Ref. \onlinecite{John's ring}), as well as
for understanding the behavior of large ensembles of interacting
rings.\cite{Davidovic 1,Davidovic 2} The statistical mechanics of these
systems requires knowledge of  ring energy levels and probabilities of
quantum or thermally activated transitions between the states of
different energies. Our work should be  useful for such statistical
modelling. Despite the known shortcomings of the London approach, this
is the only method that is of practical use for temperatures away from
the critical temperature. A key advantage of the London equations is
their linearity, which makes exact solutions for mesoscopic rings
possible.
\par

The electrostatic analogy we employed, may prove useful for various
mesoscopic sample shapes. A number of these shapes (as squares,
rectangles, or polygons) can be found in textbooks on applications of
the theory of complex functions to the two-dimensional electrostatics,
see, e.g., Ref. \onlinecite{Morse}. An example of a thin-film disk is
considered in the Appendix.
\par

\acknowledgments
We are glad to thank John Kirtley
for numerous discussions.  In part this research is supported by Grant
No. 2000011 from the United States-Israel Binational Science Foundation
(BSF), Jerusalem, Israel. This manuscript has been authored in part by
Iowa State University of Science and Technology under Contract No.
W-7405-Eng-82 with the U.S. Department of Energy.
\par
\appendix
\section{Thin film disk}

It is worth stating upfront that the solution for a disk cannot be
obtained  by setting the inner radius $a=0$ in the solutions for a
ring. The ring topology differs from that of a disk, and there is no
continuous transition from the one to another. In fact, for a disk with
no vortices the vorticity $N\equiv 0$. The state of a disk is determined
by the continuous variable $H$, the applied field, and by the vortex
position (or positions, if any).
\par

Consider a thin-film disk  of a radius $b\ll\Lambda$ with a vortex at
an arbitrary position ${\bm v}$. To find the current distribution one
has to solve Eq.(\ref{Poisson}) with the boundary condition
$G(r=b )={\rm const}=G_b$, or alternatively, Eqs. (\ref{Gv}) for $G_v$
and  (\ref{GH}) for $G_H$ under boundary conditions
$G_v(b)=0$ and $G_H(b)=G_b$.
\par

The part $G_H$ reads:
\begin{equation}
G_H(r)= \frac{cH}{8\pi\Lambda}\, r^2 + G_0\ln \frac{r}{r_0},
\label{GH(r)app}
\end{equation}
where the constants $G_0$ and $r_0$ are to be determined using the fact
that the total current is $I=-G_b$. Evaluating $I$ with the help of London
equation (\ref{London}) one should take the phase change on circles $r<v$
as zero, whereas for $r>v$ as $-2\pi$. Doing this algebra we obtain:
\begin{equation}
G_0=-\frac{c\phi_0}{4\pi^2 \Lambda}\,,\quad r_0=v\,,
\label{Go_ro}
\end{equation}
and
\begin{equation}
G_H(r)= \frac{cH}{8\pi\Lambda}\, r^2 -\frac{c\phi_0}{4\pi^2 \Lambda}  \ln
\frac{r}{v}\,.
\label{GH(r)app1}
\end{equation}
\par

The problem of finding the part $G_v$ is equivalent to one for the
electrostatic potential of a linear charge $c\phi_0/8\pi^2\Lambda$ at the
point ${\bm v}$ parallel to the grounded metallic cylinder of a radius
$b$. The solution is given, e.g., in Ref. \onlinecite{Landau}:
\begin{equation}
G_v =-\frac{c\phi_0}{4\pi^2\Lambda}
\ln\left(\frac{
b}{v}\sqrt{\frac{r^2+v^2-2rv\cos\varphi}{r^2+x^2-2rx\cos\varphi}}\,\,
\right),
\label{Gv(r,v)}
\end{equation}
where $r,\varphi$ are cylindrical coordinates, the zero azimuth is
taken as that of $\bm v$, and $x=b^2/v$. As discussed, the vortex
self-energy  $\epsilon_v=\phi_0 |G_v(\varphi=0,r\to v)|/2c$ is
logarithmically divergent, so that we set $r=v+\xi$ and $\varphi=0$ to
obtain:
\begin{equation}
\epsilon_v= \frac{\phi_0^2}{8\pi^2\Lambda} \ln \frac{b^2-v^2}{b\,\xi}.
\label{ev}
\end{equation}
The following is obtained in the same way as in the main text. The energy
in zero field is
\begin{equation}
E(0,v)=\epsilon_v(v)+ \frac{\phi_0^2}{8\pi^2\Lambda} \ln
\frac{b }{v}\,,
\label{E(0,v)}
\end{equation}
and the magnetic moment is
\begin{equation}
\mu =\frac{\phi_0(b^2-v^2)}{8\pi\Lambda} -\frac{ Hb^4}{ 16\Lambda}.
\label{mu+v_app}
\end{equation}
The thermodynamic potential ${\cal F} = E(0,v)-\int_0^H\mu\,dH$
follows:
\begin{equation}
{\cal F} = \frac{\phi_0^2}{8\pi^2\Lambda} \left(\ln
\frac{b^2-v^2}{b\,\xi}-h\,\frac{b^2-v^2}{b^2}-\frac{h^2}{4}\right)\,,
\end{equation}
where $h=\pi b^2H/\phi_0$ (note that the field normalization here
differs from the main text). In the vortex absence, ${\cal F}_0 =
(\phi_0^2/32\pi^2\Lambda)h^2$, so that the  barrier for the vortex
entry is
\begin{equation}
V_{in}^+ = \frac{\phi_0^2}{8\pi^2\Lambda} \left(\ln
\frac{b^2-v^2}{b\,\xi}-h\,\frac{b^2-v^2}{b^2}\right)\,.
\end{equation}
It is now easy to find the field at which a local minimum first appears
in the disk center ($d^2V_{in}^+/dr^2=0$ at $v=0$):
\begin{equation}
H_1 = \frac{\phi_0}{\pi b^2} \,.
\label{H*}
\end{equation}
The field $H_{c1}$ is determined by  $V_{in}^+(0)=0$:
\begin{equation}
H_{c1}=\frac{ \phi_0}{\pi b^2}\,\ln\frac{b}{ \xi}\,,
\label{Hc1_disk}
\end{equation}
the result  given by Fetter.\cite{Fetter}
\par

Thus for $H_1<H<H_{c1}$, the vortex in the disk center is in metastable
state, which becomes stable  for $H>H_{c1}$. Finally, the
barrier to the vortex entry disappears altogether when the point $v_0$
at which $V_{in}^+=0$ moves to the edge, $v_0=b-\xi$:
\begin{equation}
H^* =\frac{  \phi_0\ln 2}{2\pi\,\xi\, b}\,.
\label{Hc_disk}
\end{equation}
\par


\begin{thebibliography}{99}
%
\bibitem{Davidovic 1} D. Davidovi\`c, S. Kumar, D.H. Reich,
J. Siegel, S.B. Field, R.C. Tiberio, R. Hey, and K. Ploog, \prl {\bf
76}, 815 (1996).
%
\bibitem{Davidovic 2} D. Davidovi\`c, S. Kumar, D.H. Reich, J. Siegel,
S.B. Field, R.C. Tiberio, R. Hey, and K. Ploog, \prb {\bf 55}, 6518
(1997).
%
\bibitem{John's ring} J.R. Kirtley, C.C. Tsuei, V.G. Kogan, J.R. Clem,
H. Raffy, and Z.Z. Li, cond-mat/0302415.
%
\bibitem{Tinkham} M. Tinkham, {\it Introduction to Superconductivity},
McGrow-Hill, New York, 1996.
%
\bibitem{Buzdin} A. Bezryadin, A. Buzdin, B. Pannetier, Phys. Lett. A
{\bf 195}, 373 (1994).

\bibitem{Berger} J. Berger and J. Rubinstein,
Phil. Trans. R. Soc. Lond. A {\bf 355}, 1969 (1997).
%
\bibitem{Palacios} J.J. Palacios, \prl{\bf 84}, 1796 (2000).
%
\bibitem{Sweigert} F.M. Peters, V.A. Sweigert, B.J. Buelus, P.S. Deo,
Physica C {\bf 332}, 255 (2000).
%
\bibitem{Ioffe} L.B. Ioffe, V.B. Geshkenbein, M.V. Feigel'man,
A.L. Fauch\`ere, G. Blatter, Nature {\bf 398}, 679 (1999).
%
\bibitem{Mooij} J.E. Mooij, T.P. Orlando, L. Levitov, Lin Tian,
C.H. Van der Wal, and S. Lloyd, Science {\bf 285}, 1036 (1999).
%
\bibitem{Van der Wal} C. Van der Wal, A.C.J. ter Haar, F.K. Wilhelm,
R.N. Schouten, C.J.P.M. Harmaks, T.P. Orlando, S. Lloyd, J.E. Mooij,
Science {\bf 290}, 773 (2000).
%
\bibitem{Pearl64} J. Pearl, \apl {bf 5}, 65 (1964).
%
\bibitem{de Gennes} P.G. de Gennes, {\it Superconductivity of Metals
and Alloys}, New York, Addison-Wesley, 1989.
%
\bibitem{Fetter} A.L. Fetter, \prb {\bf 22}, 1200 (1980).
%
\bibitem{Kogan} V.G. Kogan, \prb {\bf 49}, 15874 (1994).
%
\bibitem{Morse} P.M. Morse and H. Feshbach, {\it Methods of theoretical
physics}, McGrow Hill, New York, 1953; part II, ch. 10.
%
\bibitem{Abr} {\it Handbook of Mathematical Functions}, edited by M.
Abra\-mo\-witz and A. Stegun, Natl. Bur. Stand. Appl. Math. Ser. No. 55
(U.S. GPO, Washington, D.C., 1965).
%
\bibitem{rem1} It is instructive to compare the vortex energy at the
ring, Eq.\ (\ref{e_v}) with an expression for a narrow straight strip:
$\epsilon_v = (\phi_0^2/8\pi^2\Lambda) \ln [(2W/\pi\xi)\sin(\pi
v/W)]$, where $W\ll \Lambda$ is the strip width and $v$ is the vortex
distance from one of the edges.\cite{Kogan} One should mention that the
general result for a vortex near the edge of a half-infinite thin film
of Ref.~\onlinecite{Kogan} is erroneous; still, the discussion of
energies and of the samples small on the scale of $\Lambda$ is correct.
%
\bibitem{Landau} L.D. Landau and E.M. Lifshitz, {\it Electrodynamics of
continuous media}, Pergamon, Oxford, New York, 1984.
%
\bibitem{Barone} A. Barone and G. Paterno, {\it Physics and Applications
of the Josephson Effect}, Wiley, New York, 1982, p. 355.
%
\bibitem{Clemunpub} J.R. Clem, unpublished.
%
\bibitem{Dantsker97}
E. Dantsker, S. Tanaka, and J. Clarke, \apl {\bf 70}, 2037 (1997).
%
\end{thebibliography}
\end{document}